\documentclass[a4paper,11pt]{article}
\usepackage[dvips]{graphicx}
\usepackage{amssymb}
\usepackage{amsmath}

\usepackage{cite}

\begin{document}

\title{A Note on Gravitational Memory in F(R)-theories and their Equivalent Scalar-Tensor Theories}
\author{
V.K. Oikonomou\,\thanks{voiko@physics.auth.gr}\\
Department of Theoretical Physics, Aristotle University of Thessaloniki,\\
54124 Thessaloniki, Greece
} \maketitle

\begin{abstract}
In this paper we consider the implications that the effect gravitational memory would have on primordial black holes, within the theoretical context of $F(R)$ related scalar-tensor theories. As we will demonstrate, under the assumption that the initial mass of the primordial black hole is such so that it evaporates today, this can potentially constrain the $F(R)$ related theories of gravity. We study two scalar-tensor models and discuss the evolution of primordial black holes created at some initial time $t_f$ in the early universe. The results between the two models vary significantly which shows us that, if the effect of gravitational memory is considered valid, some of the scalar-tensor models and their corresponding $F(R)$ theories must be further constrained.  
\end{abstract}

\section*{Introduction}

General Relativity is the most successful theory for describing nature at scales in which gravitational interactions become important. Therefore, it is the best theoretical tool to describe local gravitational interactions of massive objects and in addition to describe how the universe evolves as a whole. One of the most striking astrophysical observations that utterly changed our perspectives about the cosmological evolution came in the late 90's and according to these, the universe has undergone two accelerating phases. The first accelerating phase is the inflation period and the second is the present epoch's acceleration, which is referred to as dark energy. The new Planck telescope observational data corresponding to the present epoch indicate that the universe is described by the $\mathrm{\Lambda}\mathrm{CDM}$ model with the main features of this model being that the universe is nearly spatially flat, and consists of ordinary matter ($\sim4.9\%$), cold dark matter ($\sim 26.8\%$) and dark energy ($\sim 68.3\%$).

\noindent A very theoretically consistent and promising description of the dark energy and the late time acceleration this generates, is provided by the so called $F(R)$ modified theories of gravity. For informative reviews and very important papers on these theories, consult \cite{reviews,importantpapers} and references therein. In these theories what changes drastically is the right hand side of the Einstein equations. In order to obtain accelerating solutions of the Friedmann-Robertson-Walker equations, the energy momentum tensor necessarily must contain a negative $w$ fluid, a feature naturally provided by these $F(R)$ theories. More importantly and surprisingly, early time inflation and late time acceleration may successfully described by $F(R)$ theories \cite{reviews,importantpapers,sergeinojirimodel,generalfrtheories,singularitiesaccelerationmodels,exponentialmodels}. The possibility to unify early-time inflation with late-time acceleration in $F(R)$ gravity was explicitly demonstrated in the Nojiri-Odintsov model \cite{sergeinojirimodel}. Moreover, exact solutions in strong gravitational background have been presented in \cite{solutions}. In addition, an important stream of papers on modified gravitational theories with non-minimal curvature matter coupling can be found in \cite{bertolami,bertolami1}. Notice that in the present paper when we use the expression non-minimal coupling, we mean the term containing the scalar field coupled to the curvature, which can be generally of the form $f(\varphi )R$, with $R$ the curvature scalar and $\varphi $ the scalar field \cite{faraoni,fujii}. In this fashion, even the Brans-Dicke model is a non-minimally coupled scalar-tensor theory \cite{faraoni,fujii}, and also assume that there exists no direct coupling between matter and the curvature scalar.

Since $F(R)$ theories serve as consistent generalizations of general relativity, these theories are confronted with the successes of general relativity and a large number of constraints need to be satisfied, in order a modified gravity theory can be considered as viable. Most of these constraints are imposed from local tests of general relativity, coming from planetary and star formation tests and additionally from various cosmological bounds. Moreover, since each Jordan frame $F(R)$ theory has a scalar-tensor gravitational theory counterpart in the Einstein frame, the scalarons of this counterpart theory have to be classical, so that quantum-mechanical stability is ensured.

With regard to the constraints imposed on $F(R)$ theory, in this article we shall indicate another source of constraints that can be imposed to $F(R)$ theories to ensure their viability. These constraints come from Primordial Black Holes and their evolution in a flat Friedmann-Robertson-Walker (FRW hereafter) cosmological background. 

Primordial Black Holes (PBH hereafter) are light black holes that can be formed in the early universe, probably after the inflationary period of the universe ends (for a recent review see \cite{reviewgreen}). These black holes can be the result of various formation ways, for example collapsing of large density perturbations, cosmic string loops and bubble collisions. Since the evaporation of PBH is inevitable (if these radiate particles in the normal way), there exist a number of constraints on these objects. Moreover, since some formation mechanisms include large density perturbations, the PBH power spectrum of density perturbations on small scales can be used to constrain models of inflation. In addition, since PBH are formed in a time before nucleosynthesis, PBH can serve for a potential cold dark matter candidate. 

The focus in this paper is on specific evolution aspects of PBH, from the point of view of $F(R)$ related scalar-tensor theories. We shall assume that the PBH have an evaporation time equal to the present age of the universe, discussing only the other possibility that the PBH do not have evaporation time equal to the present age of the universe. The evaporation time determines the initial mass of the PBH, which depends on the gravitational constant at the time of formation, if gravitational memory takes place \cite{barrow}. We shall study the evolution of PBH, using two scalar-tensor models, adopting the gravitational memory effect introduced by Barrow \cite{barrow}, which states that the PBH carries information about the value of the gravitational constant at the time of its formation. This can potentially constrain the viable $F(R)$  models that exist in the literature, assuming always that the gravitational memory effect is valid. In addition, since in $F(R)$ theories the anti-evaporation of black holes is possible \cite{odintsovantievaporation}, we briefly discuss the possibility that a PBH develops to be a supermassive black hole during the evolution process of the universe.

This article is organized as follows: In section 1, we present the general features of $F(R)$ theories in the Jordan frame in the metric formalism. In section 2, we introduce a general class of scalar-tensor theories we shall use and demonstrate their equivalence to $F(R)$ theories using a specific Weyl transformation. The two scalar-tensor models predict late-time acceleration. In addition, we study the evolution of the gravitational constant for both the scalar-tensor models and show that the results may vary significantly. We also analyze the evolution of PBH using these two solutions coming from the scalar-tensor models. Moreover, we compare the results with two Brans-Dicke models. As we shall see, one of the two scalar-tensor models has un-realistic implications for the PBH and therefore this result should be critically taken into account. In section 3, we briefly discuss the possibility that PBH, if these are linked to $F(R)$ theories of gravity, can develop to be supermassive black holes. This is possible since anti-evaporating black holes occur in $F(R)$ theories. Some antigravity issues are briefly discussed in the end of section 3 and the conclusions follow in the end of the article.

\section{General Features of $F(R)$ Dark Energy Models in the Jordan Frame}

Before getting into the core of the article's subject, it worths giving a brief description of the main features of $F(R)$ gravity theories in the Jordan frame. For an important stream of reviews and articles see \cite{reviews,importantpapers} and references therein. The $F(R)$ theories of modified gravity are described by the following four dimensional action, which is a direct modification of the Einstein-Hilbert action:
\begin{equation}\label{action}
\mathcal{S}=\frac{1}{2\kappa^2}\int \mathrm{d}^4x\sqrt{-g}F(R)+S_m(g_{\mu \nu},\Psi_m),
\end{equation}
with $\kappa^2=8\pi G$ and $S_m$ the matter action of the matter fields $\Psi_m$. We shall assume that the form of the $F(R)$ theory that we shall use is of the form $F(R)=R+f(R)$ and also we shall work in the metric formalism. Within the metric formalism context, by varying the action (\ref{action}) with respect to $g_{\mu \nu}$, we get the equations of motion:
\begin{equation}\label{eqnmotion}
F'(R)R_{\mu \nu}(g)-\frac{1}{2}F(R)g_{\mu \nu}-\nabla_{\mu}\nabla_{\nu}F'(R)+g_{\mu \nu}\square F'(R)=\kappa^2T_{\mu \nu}.
\end{equation} 
In the above, $F'(R)=\partial F(R)/\partial R$ and $T_{\mu \nu}$ stands for the energy momentum tensor. 

The underlying theme and the most important feature of the $F(R)$ modified gravity theories is that, the right hand side of the Einstein equations are modified and not the left. Indeed, we can write the above equations of motion in the following form:
\begin{align}\label{modifiedeinsteineqns}
R_{\mu \nu}-\frac{1}{2}Rg_{\mu \nu}=\frac{\kappa^2}{F'(R)}\Big{(}T_{\mu \nu}+\frac{1}{\kappa}\Big{[}\frac{F(R)-RF'(R)}{2}g_{\mu \nu}+\nabla_{\mu}\nabla_{\nu}F'(R)-g_{\mu \nu}\square F'(R)\Big{]}\Big{)}.
\end{align}
Thereby, the energy momentum tensor acquires an additional contribution coming from the term:
\begin{equation}\label{newenrgymom}
T^{eff}_{\mu \nu}=\frac{1}{\kappa}\Big{[}\frac{F(R)-RF'(R)}{2}g_{\mu \nu}+\nabla_{\mu}\nabla_{\nu}F'(R)-g_{\mu \nu}\square F'(R)\Big{]},
\end{equation}
and this term is what actually models the dark energy in these modified gravity theories. By taking the trace of equation (\ref{eqnmotion}) we obtain the following equation:
\begin{equation}\label{traceeqn}
3\square F'(R)+R F'(R)-2F(R)=\kappa^2 T,
\end{equation}
with $T$ standing for the trace of the energy momentum tensor $T=g^{\mu \nu}T_{\mu \nu}=-\rho+3P$, while $\rho$, $P$ stand for the energy density and pressure of the matter respectively. Equation (\ref{traceeqn}) actually reveals another degree of freedom of the $F(R)$ theories, governed by the scalar field $F'(R)$. Actually, equation (\ref{traceeqn}) is the equation of motion of this scalar degree of freedom, which is nothing but the ``scalaron'' field. Considering a flat Friedmann-Lemaitre-Robertson-Walker spacetime, the Ricci scalar is given by:
\begin{equation}\label{ricciscal}
R=6(2H^2+\dot{H}),
\end{equation}
with $H$ denoting the Hubble parameter, and the ``dot'' indicates differentiation with respect to time. The cosmological dynamics are governed by the following equations:
\begin{align}\label{flrw}
& 3F'(R)H^2=\kappa^2(\rho_m+\rho_r)+\frac{(F'(R)R-F(R))}{2}-3H\dot{F}'(R), \\ \notag &
-2F'(R)\dot{H}=\kappa^2(p_m+4/3\rho_r)+F\ddot{F}'(R)-H\dot{F}'(R),
\end{align}
where $\rho_r$ and $\rho_m$ stand for the radiation and matter energy density respectively. Therefore, the total effective energy density and pressure of matter and geometry are \cite{reviews}:
\begin{align}\label{densitypressure}
& \rho_{eff}=\frac{1}{F'(R)}\Big{[}\rho+\frac{1}{\kappa^2}\Big{(}F'(R)R-F(R)-6H\dot{F}'(R)\Big{)}\Big{]} \\ \notag &
p_{eff}=\frac{1}{F'(R)}\Big{[}p+\frac{1}{\kappa^2}\Big{(}-F'(R)R+F(R)+4H\dot{F}'(R)+2\ddot{F}'(R)\Big{)}\Big{]},
\end{align} 
with $\rho,P$ standing for the total matter energy density and pressure respectively.

\section{F(R) theories, Scalar-Tensor Theories and Study PBH Using two scalar-tensor Models}

Consider the following action describing a non-minimal coupled scalar field to the Ricci scalar $R$, in $D-$dimensions:
\begin{equation}\label{generaleqn}
S=\int \mathrm{d}x^D\sqrt{-g}\Big{[}(1+f(\phi))R-\frac{1}{2}\omega (\phi)\partial_{\mu}\phi\partial^{\mu}\phi-V(\phi )\Big{]}
\end{equation}
This type of scalar-tensor models was extensively studied in reference \cite{odintsov1}, for the four dimensional case and are known to produce late-time acceleration in both the Einstein and Jordan frames \cite{odintsov1}. Notice that the effective gravitational constant in the Jordan frame for this type of theories is:
\begin{equation}\label{effgravcnst}
G_{eff}=\frac{1}{16\pi (1+f(\phi ))}
\end{equation}

\subsection{A general technique to obtain an $F(R)$ theory from a non-minimally coupled scalar-tensor theory}

Let us start our analysis by demonstrating that the scalar-tensor theories described by the action (\ref{generaleqn}) can be equivalent to $F(R)$ theories. Usually in the literature one actually starts from an $F(R)$ theory and ends up to a non-minimally coupled scalar-tensor theory, which is actually a Brans-Dicke theory with $\omega_{BD}$ equal to zero (see \cite{reviews} and particularly Odintsov and Nojiri (2007) and Felice, Tsujikawa (2010)). A similar method to the one we shall use appears in \cite{odintsov2}, but applied in four dimensions. Before going into the details, it worths recalling how the equivalence between metric $F(R)$ theories and scalar-tensor theories with non-minimal coupling (Brans-Dicke theories with $\omega_{BD}=0$) is established in the literature. Following \cite{reviews}, consider the general $F(R)$ theory with matter, which is described by the action:
\begin{equation}\label{actionlagmult}
\mathcal{S}=\frac{1}{2\kappa^2}\int \mathrm{d}^4x\sqrt{-g}F(R)+S_m(g_{\mu \nu},\Psi_m)
\end{equation}
Using the Lagrange multipliers method, we introduce an auxiliary field $\chi$ which acts as a Lagrange multiplier, and so the action (\ref{actionlagmult}) becomes:
\begin{equation}\label{actionlagmult1}
\mathcal{S}=\frac{1}{2\kappa^2}\int \mathrm{d}^4x\sqrt{-g}\Big{(}F(\chi)+F_{,\chi}(\chi)(R-\chi)\Big{)}+S_m(g_{\mu \nu},\Psi_m)
\end{equation}
with $F_{,\chi}(\chi)$ standing for the first derivative of the function $F(\chi)$ with respect to $\chi$. Varying the action (\ref{actionlagmult1}) with respect to $\chi$ one obtains:
\begin{equation}\label{fgegdffdbkf}
F_{,\chi \chi}(\chi)(R-\chi)=0
\end{equation}
Given that $F_{,\chi \chi}(\chi)\neq 0$, which is true for most interesting $F(R)$ theories, we may conclude that $R=\chi$. Therefore, the action (\ref{actionlagmult1}) recovers the initial $F(R)$-gravity action (\ref{actionlagmult1}). Defining,
\begin{equation}\label{newdef}
\varphi =F_{,\chi}(\chi),
\end{equation}
the action of equation (\ref{actionlagmult1}) can be expressed as a function of the field $\varphi$ as follows:
\begin{equation}\label{actionlagmult123}
\mathcal{S}=\int \mathrm{d}^4x\sqrt{-g}\Big{[}\frac{1}{2\kappa^2}\varphi R-U(\varphi )\Big{]}+S_m(g_{\mu \nu},\Psi_m)
\end{equation}
with the potential $U(\varphi )$ being equal to:
\begin{equation}\label{potentialbdfr}
U(\varphi )=\frac{\chi (\varphi)\varphi -F(\chi (\varphi ))}{2\kappa^2}
\end{equation}
The function $\chi (\varphi)$ can be obtained by solving the algebraic equation (\ref{newdef}) with respect to $\chi $, so that $\chi$ is a function of $\varphi$. From equation (\ref{actionlagmult123}), it easily follows that the $F(R)$ theories in the metric formalism are equivalent to Brans-Dicke theories with the parameter $\omega_{BD}$ equal to zero. Therefore the metric $F(R)$ theories are equivalent to non-minimally coupled scalar-tensor theories with $\omega =0$. Mark this latter constraint, that is, the disappearance of the kinetic term of the scalar field, because it will appear as a constraint in the method we shall use. 

In the method we shall present, we shall mainly be interested in metrics that belong to the conformal structure of the pseudo-Riemannian manifold under study. We shall take into account two metrics which are conformally related to each other. This conformal relation is materialized by a Weyl rescaling of one of the two metrics, which we shall assume to be of the following form,
\begin{equation}\label{weylrescale}
\tilde{g}_{\mu \nu}=e^{2\eta (\phi )}g_{\mu \nu}
\end{equation}
with $\eta (\phi )$ an arbitrary real function of the scalar field $\phi$ and where it appears $\Omega =e^{2\eta (\phi )}$. Note that $g_{\mu \nu}=e^{-2\eta (\phi )\tilde{g}_{\mu \nu}}$. In the same way $\sqrt{-\tilde{g}}=\Omega^{-D}\sqrt{-g}$. The manifold we shall assume as a geometrical background is a pseudo-Riemannian one, which is described by a Lorentz metric (which in most cases is the FRW metric) and a symmetric (torsion-less) compatible with the metric affine connection, the Levi-Civita connection. The Christoffel symbols in such a geometric background is: 
\begin{equation}\label{christofell}
\Gamma_{\mu \nu }^k=\frac{1}{2}g^{k\lambda }(\partial_{\mu }g_{\lambda \nu}+\partial_{\nu }g_{\lambda \mu}-\partial_{\lambda }g_{\mu \nu})
\end{equation} 
and additionally, the Ricci scalar is equal to:
\begin{equation}\label{ricciscalar}
R=g^{\mu \nu }(\partial_{\lambda }\Gamma_{\mu \nu }^{\lambda}-\partial_{\nu }\Gamma_{\mu \rho }^{\rho}-\Gamma_{\sigma \nu }^{\sigma}\Gamma_{\mu \lambda }^{\sigma}+\Gamma_{\mu \rho }^{\rho}g^{\mu \nu}\Gamma_{\mu \nu }^{\sigma})
\end{equation} 
Under the Weyl rescaling (\ref{weylrescale}), the Ricci scalar term appearing in the action (\ref{generaleqn}) is transformed in the following way:
\begin{align}\label{weyltransformed}
R=\Omega^2(1+f(\phi ))\tilde{R}+\Omega^2(2(D-1)\tilde{\square } f(\phi )-(D-2)(D-1)\tilde{g}^{\mu \nu }\partial_{\mu}f(\phi )\partial_{\nu}f(\phi )\end{align}
with,
\begin{equation}\label{fdssdgdsg}
\partial_{\mu}f(\phi )=\partial_{\mu}\Omega(\phi )
\end{equation}
and in addition,
\begin{equation}\label{fdssdgdsg}
\tilde{\square }f(\phi )=\frac{1}{\sqrt{-\tilde{g}}}\partial_{\mu}\Big{(}\sqrt{-\tilde{g}}\partial^{\mu}\ln \Omega (\phi)\Big{)}
\end{equation}
Accordingly, the action of relation (\ref{generaleqn}) now becomes:
\begin{align}\label{generaleqn1}
S=&\int \mathrm{d}x^D\sqrt{-\tilde{g}}\Big{[}\big{(}1+f(\phi)\big{)}\Big{(}\Omega^2(\phi)\tilde{R}-\Omega^2(\phi)(D-1)(D-2)\partial_{\mu}\phi\partial^{\mu}\phi\eta'(\phi)\eta'(\phi)\Big{)}
\\ \notag &-\frac{1}{2}\omega (\phi)\Omega^2(\phi)\partial_{\mu}\phi\partial^{\mu}\phi-\Omega^4(\phi)V(\phi )\Big{]}
\end{align}
where we assumed that the surface term $\int \mathrm{d}x^{D-1}\sqrt{-\tilde{g}}\partial_{\mu}\phi \eta'(\phi)\rightarrow 0$. The action (\ref{generaleqn1}) can be re-written as follows:
\begin{align}\label{generaleqn2}
S=&\int \mathrm{d}x^D\sqrt{-\tilde{g}}\Big{[}(1+f(\phi))e^{2\eta (\phi)}\tilde{R}\\ \notag &-(1+f(\phi))e^{2\eta (\phi)}(D-1)(D-2)\partial_{\mu}\phi\partial^{\mu}\phi\eta'(\phi)\eta'(\phi)
 -\frac{1}{2}\omega (\phi)e^{2\eta (\phi)}\partial_{\mu}\phi\partial^{\mu}\phi-e^{4\eta (\phi)}V(\phi )\Big{]}
\end{align}
Now the equivalence of action (\ref{generaleqn2}) to $F(R)$ theories depends crucially on the following choice, 
\begin{equation}\label{algbreqn}
 -\big{(}1+f(\phi)\big{)}(D-1)(D-2)[\eta'(\phi)]^2+\frac{\omega(\phi)}{2}=0 
\end{equation}
If (\ref{algbreqn}) holds true, then the action (\ref{generaleqn2}) can be rewritten as follows:
\begin{align}\label{generaleqn3}
S=&\int \mathrm{d}x^D\sqrt{-\tilde{g}}\Big{[}(1+f(\phi))e^{2\eta (\phi)}\tilde{R}-e^{4\eta (\phi)}V(\phi )\Big{]}
\end{align}
The constraint (\ref{algbreqn}) makes the kinetic term of relation (\ref{generaleqn2}) disappear, that is, the term analogous to $\partial_{\mu}\phi\partial^{\mu}\phi$. Recall that on going from $F(R)$ theories to their equivalent scalar-tensor theories, we also end up to a scalar-tensor theory with the kinetic term of the scalar field being equal to zero ($\omega_{BD}=0$).

Now we may obtain an $F(R)$ theory in a straightforward way, by inverting $\phi=\phi (\tilde{R})$ using equation (\ref{algbreqn}) and solving with respect to $\phi$. Then, the corresponding $F(\tilde{R})$ theory is described by the action:
\begin{align}\label{generaleqn4}
S=&\int \mathrm{d}x^D\sqrt{-\tilde{g}}F(\tilde{R}),
\end{align}
with $F(\tilde{R})$ standing for:
\begin{equation}\label{frequation}
F(\tilde{R})=(1+f(\phi (\tilde{R})))e^{2\eta (\phi (\tilde{R}))}\tilde{R}-e^{4\eta (\phi (\tilde{R}))}V(\phi (\tilde{R}))
\end{equation}
So having established the equivalence under certain assumptions (which actually is relation (\ref{algbreqn}) and the vanishing of the surface term we saw previously) between a general scalar-tensor theory of type (\ref{generaleqn}) and $F(R)$ theories, let us examine two scalar-tensor models with appealing attributes regarding their cosmological implications in both the Einstein and Jordan frames. Particularly, we shall focus on the impact of these solutions to the gravitational memory effect in PBH containing the phantom matter of the form (\ref{generaleqn}) in dimensions four.  

Before closing this subsection, we have to mention that we assumed the $F(R)$ theory has no direct coupling to the matter Lagrangian. However, there exists in the literature the particularly interesting case in which the $F(R)$ theory is directly coupled to the matter Lagrangian \cite{bertolami}, leading to new interesting phenomena, such as the prediction of a fifth force \cite{bertolami}. We defer in a future work the study of such $F(R)$ theories in the context of gravitational memory. In addition, a study of the same phenomena in the Palatini formalism also deserves some attention which are beyond the scope of this article.

A final comment regarding the technique we used. One may naturally ask why such an essential ansatz described by relation (\ref{algbreqn}) is needed in order to establish an equivalence between a general non-minimally coupled scalar-tensor theory (not just a Brans-Dicke model with a potential and $\omega_{BD}=0$) and an $F(R)$ theory. The answer to this was given partially in the above text but it worths to answer this issue in order to make things much more clear. In principle our technique describes that every non-minimally coupled scalar-tensor theory can have a specific $F(R)$ counterpart. This $F(R)$ theory has to satisfy equation (\ref{algbreqn}), so its not a general $F(R)$ theory, but belongs to some restricted class of $F(R)$ theories. Note that there is not a direct coupling to matter, at least in the Lagrangian level, which is conceptually different from the $F(R)$ theories that were studied in \cite{bertolami}. The constraint (\ref{algbreqn}) is very essential and it is conceptually similar to the requirement that the final $F(R)$ theory is in some way related to a scalar-tensor theory without kinetic term for the scalar field, which is described by the conformally transformed action (\ref{generaleqn2}) (something that happens naturally when one goes from $F(R)$ theories to scalar-tensor theories as we described earlier). A similar ansatz was also used by the authors of \cite{odintsov2}, but in a different context. We believe this constraint is the link between the two completely different methods describing the two ways from going from scalar-tensor theories to $F(R)$ theories and back. Indeed, if we start from a general $F(R)$ theory, we end up to a non-minimally coupled scalar-tensor theory with vanishing kinetic term of the scalar field and with a potential term of the scalar field (a Brans-Dicke theory with $\omega_{DB}=0$ and a potential term). Action (\ref{generaleqn2}) describes a non-minimally coupled scalar-tensor theory with vanishing kinetic term (assuming that equation (\ref{algbreqn}) holds true) which, with the aid of the constraint (\ref{algbreqn}), can be equivalent to an $F(R)$ theory. Moreover action (\ref{generaleqn2}) is conformally related to a non-minimally coupled scalar-tensor theory with non-vanishing kinetic term described by action (\ref{generaleqn}).

\subsection{General features of gravitational memory and PBH}

Gravitational memory of black holes \cite{barrow1} is an effect according to which each black hole is considered as a small gravitational structure in an expanding universe. This means that a PBH retains information about the time it was formed, since the gravitational constant inside the PBH remains the one it was at the time of its formation, but outside the black hole the gravitational constant evolves with time. Suppose that the PBH was formed at a time $t_f$ and let the gravitational constant at the time of its formation be $G(t_f)$. Then the evolution of each PBH can be significantly altered depending on its initial mass. Since the PBH are losing energy by Hawking radiation, which is directly affected by gravitational memory, this effect can have significant observational implications which strongly depend on the initial mass of the PBH and on the $G(t_f)$ value. There are three possibilities for the initial mass $M_i$ of the PBH:
\begin{itemize}

\item $M_i<M_{*}$

\item $M_i=M_{*}$

\item $M_i>M_{*}$

\end{itemize}
where $M_{*}$ is the critical initial mass of PBH which are considered to be completely evaporated at present time, so the evaporation time is equal to the present age of the universe. We shall study the PBH that are exactly of this type, so they have initial mass equal to $M_{*}$. We shall discuss possible constraints that the other two cases can impose on $F(R)$ theories, in the end of this section. Of course the case in which $M_i=M_{*}$ can also constrain some $F(R)$ theories by allowing or not their scalar-tensor counterparts. We shall briefly discuss these constraints in the end of this section.

Having assumed that the PBH has a total evaporation time equal to the present age of the universe, let us see what this implies for the initial mass of a PBH. The initial mass of a PBH is of the order of the particle horizon:
\begin{equation}\label{initialmass}
M_i\simeq G^{-1}t
\end{equation}
The evaporation procedure follows the Hawking radiation law, according to which the PBH emits particles like a blackbody with rate \cite{barrow1}:
\begin{equation}\label{gfjdhf}
-\frac{\mathrm{d}M}{\mathrm{d}t}=af(M)M^{-2}\sim T(M)^4
\end{equation}
with $f(M)$ the number of particle species which can be emitted, that is, with rest mass below $T(M)=(8\pi G M)^{-1}$. The total evaporation time for a initial mass $M_i$ PBH is:
\begin{equation}\label{lft}
\tau_{BH}=aG^2(t_f)M_i^3
\end{equation}
Recall that $t_f$ is the time when the PBH was formed and notice that the gravitational constant at that time $G(t_f)$, determines both the Hawking temperature and the lifetime of PBH. Now, in our case for which the lifetime equals the present age of the universe $t_0$, then the initial mass is the critical mass $M_*$:
\begin{equation}\label{inmasscrit}
M_*=\Big{(}\frac{t_0}{aG^2(t_0)}\Big{)}^{1/3}\Big{(}\frac{G(t_0))}{G(t_f)}\Big{)}^{2/3}
\end{equation}
Accordingly, the Hawking temperature for such critical mass PBH is:
\begin{equation}\label{hawktmpre}
T_{BH}=A\Big{(}\frac{G(t_0))}{G(t_f)}\Big{)}^{1/3}
\end{equation}
Assume that the time at which the PBH were formed is $t_f=10^{-21}$ (the same as used in Barrow \cite{barrow}), after the inflationary period (started at $t=10^{-35}$sec and ended at $t=10^{-33}$) has ended and before baryogenesis. Of course we may vary the initial time of formation but we are not so interested to exact phenomenology but just to have a solid idea of the general picture.

Now let us see the implications of two scalar-tensor models of the type (\ref{generaleqn1}) in dimensions four, on the gravitational memory effect and compare the results of these to the ones obtained from a Brans-Dicke model with and without cosmological constant. Note that both the models we shall present are equivalent to an $F(R)$ theory. We use the models studied in reference \cite{odintsov1}. Before going into the details of the two models, let us note that the models we shall study have a time varying gravitational constant that can have direct impact on observational data of PBH. Therefore, the gravitational memory effect can in some way constrain $F(R)$ theories via constraining their equivalent scalar-tensor theories. Note however that the models we shall use are just for exposition purposes only and we should consistently add matter to these in order to get a detailed phenomenology.

\subsection{Scalar-Tensor Model 1}

A model that leads to late time acceleration, that is, acceleration that has recently started in the redshift scale, was studied in reference \cite{odintsov1}. The model is described by the Jordan frame action appearing in equation (\ref{generaleqn}), in four dimensions, with the function $f(\phi)$ being equal to:
\begin{equation}\label{fder}
f(\phi)=\frac{1-a_1\phi}{a_1\phi}
\end{equation}
The kinetic function $\omega (\phi)$ and the potential $V(\phi )$, appearing (\ref{generaleqn}), for the function (\ref{fder}) are equal to:
\begin{equation}\label{kintpotent}
\omega (\phi)=-\frac{3}{a_1^2\phi^3},{\,}{\,}{\,}V(\phi )=\frac{6\tilde{H_0}}{a_1^2\phi^2}
\end{equation}
Note that in these kind of models described by relation (\ref{generaleqn1}) in dimensions four, the energy density and the pressure of the scalar field $\phi$ are equal to:
\begin{equation}\label{enrgydensitypressure}
\rho_{\phi}=\frac{1}{2}W(\phi)\dot{\phi}^2+U(\phi ),{\,}{\,}{\,}p_{\phi}=\frac{1}{2}W(\phi)\dot{\phi}^2-U(\phi )
\end{equation}
with the function $W(\phi )$ being equal to:
\begin{equation}\label{wufunctions}
W(\phi )=\frac{\omega (\phi)}{1+f(\phi )}+\frac{3}{(f(\phi )+1)^2}\Big{(}\frac{\mathrm{d}f(\phi )}{\mathrm{d}\phi}\Big{)}^2
\end{equation}
and also the function $U(\phi )$ is equal to:
\begin{equation}\label{wufunctions}
W(\phi )=\frac{V (\phi)}{(f(\phi )+1)^2}
\end{equation}
When the model is considered in the Jordan frame, the effective gravitational constant is time-dependent and varies as the field $\phi (t)$ varies with time. Particularly, in the case of the function (\ref{fder}) the effective gravitational constant is equal to (see equation (\ref{effgravcnst})):
\begin{equation}\label{effgr1}
G_{eff}=\frac{a_1\phi (t)}{16\pi}
\end{equation}
Assuming a spatially flat FRW metric of the form,
\begin{equation}\label{metricformfrw}
\mathrm{d}s^2=-\mathrm{d}t^2+a^2(t)\sum_i\mathrm{d}x_i^2
\end{equation}
with $a(t)$ the scale factor, we may solve the cosmological equations of motion, with the solution $\phi (t)$ corresponding to the model (\ref{fder}) in the Jordan frame being equal to \cite{odintsov1}:
\begin{equation}\label{phit1}
\phi (t)=\frac{1}{a_1}\Big{(}\frac{3a_1t}{2}\Big{)}^{2/3}
\end{equation}
Note that the specific model has the particularly interesting attribute of giving late time acceleration in both the Einstein frame and in the Jordan frame. We shall be mostly interested in the Jordan frame solution, in which the gravitational constant varies with respect to time as follows:
\begin{equation}\label{fhf}
G_{eff}(t)=\frac{1}{16\pi a_1}\Big{(}\frac{3a_1t}{2}\Big{)}^{2/3}
\end{equation}
with $a_1$ a model dependent constant. Now suppose that the universe is filled with this scalar field at the time of PBH formation and suppose the PBH was formed at some time $t_f=10^{-21}$sec after the Big Bang. In what follows we shall use the fraction $G_{eff}(t_f)/G_{eff}(t_0)$, with $t_0$ the present time. We shall assume that the PBH have formed according to the gravitational memory effect we described earlier and in addition, that the PBH have mass which is equal to the critical mass (\ref{inmasscrit}) and also the Hawking temperature at present time is given by (\ref{hawktmpre}). What is actually severely changed in the PBH evolution, in the context of gravitational memory, is the temperature and mass of PBH which undergo a complete evaporation today. Notice that the appearance of the fraction $G_{eff}(t_f)/G_{eff}(t_0)$ in both relations (\ref{inmasscrit}) and (\ref{hawktmpre}), may severely alter the present day observable effects of PBH emission of particles. It is therefore crucial for the present model to evaluate the fraction $G_{eff}(t_f)/G_{eff}(t_0)$, which is equal to:
\begin{equation}\label{effec}
\frac{G_{eff}(t_f)}{G_{eff}(t_0)}=\frac{1}{16\pi a_1}\Big{(}\frac{3a_1t}{2}\Big{)}^{2/3}\frac{1}{G_{eff}(t_0)}
\end{equation}
Using the present value of $G_{eff}(t_0)$, and for $t_f=10^{-21}$sec, the above fraction is approximately equal to: 
\begin{equation}\label{apprxvl}
\frac{G_{eff}(t_f)}{G_{eff}(t_0)}\simeq 3.8\times 10^{-6}
\end{equation}
for the present model. Of course this can vary since it depends on $a_1$, which we took equal to one. Nevertheless, we have an idea of the order of the fraction. Let us see what this implies for the PBH particle emission in present time. According to Barrow, and it can be easily verified from relations (\ref{inmasscrit}) and (\ref{hawktmpre}), for $G_{eff}(t_f)/G_{eff}(t_0) \sim 3.8\times 10^{-6}$, the evaporation Hawking temperature of the PBH is approximately equal to $T_{BH}\simeq 0.37$MeV, which is below the electron rest mass and therefore the possibility of observing black hole explosions via the observation of Gamma ray and radio bursts created by relativistic electron and positrons \cite{barrow}. This result can be modified depending the initial value of the constant $a_1$, which is model dependent. The value of this constant may determine the late time acceleration of the scalar-tensor model. For example a Hawking temperature in the KeV scale and the photon emission from PBH is mainly in the X-ray band and massive particles would not be emitted (apart from light mass particles).

Finally, let us note that, as we evinced earlier, the scalar-tensor theory described by the function (\ref{fder}), corresponds to an $F(R)$ theory which can be found by making use of relations (\ref{algbreqn}) and (\ref{frequation}). Therefore, we may impose some indirect constraints on the $F(R)$ theory by constraining the corresponding scalar-tensor theory using gravitational memory and present time black hole evaporation data in various bands. The complete study of various $F(R)$ models by using gravitational memory of PBH will be reported in a future work.

\subsection{Scalar-Tensor Model 2}

The second scalar-tensor model we shall study also comes from reference \cite{odintsov1} and is described by the Jordan frame action appearing in equation (\ref{generaleqn}), in four dimensions, with the function $f(\phi )$ this time being equal: 
\begin{equation}\label{fder1}
f(\phi)=\phi-t_0
\end{equation}
Then, the kinetic function $\omega (\phi)$ and the potential $V(\phi )$, are equal to:
\begin{equation}\label{kintpotent1}
\omega (\phi)=-\frac{3}{\phi+1-t_0},{\,}{\,}{\,}V(\phi )=6\tilde{H_0}(1+\phi-t_0)
\end{equation}
Therefore, the effective time-dependent gravitational constant is equal to (see equation (\ref{effgravcnst})):
\begin{equation}\label{effgr11}
G_{eff}=\frac{1}{16\pi (\phi (t)-t_0+1)}
\end{equation}
Again assuming a spatially flat FRW metric of the form (\ref{metricformfrw}), the solution $\phi (t)$ corresponding to the model (\ref{fder1}) in the Jordan frame is \cite{odintsov1}:
\begin{equation}\label{phit11}
\phi (t)=\frac{t^2}{4}+t_0-1
\end{equation}
and thereby the gravitational constant varies with respect to time as follows:
\begin{equation}\label{fhf1}
G_{eff}(t)=\frac{1}{4\pi t^2}
\end{equation}
We can easily calculate the fraction $G_{eff}(t_f)/G_{eff}(t_0)$, which in this case is equal to:
\begin{equation}\label{effec1}
\frac{G_{eff}(t_f)}{G_{eff}(t_0)}=\frac{1}{4\pi t^2}\frac{1}{G_{eff}(t_0)}
\end{equation}
Giving the same numerical values to the parameters as we did in the previous model, we obtain an estimate for the value of the fraction $G_{eff}(t_f)/G_{eff}(t_0)$, which is approximately equal to:
\begin{equation}\label{apprxvl2}
\frac{G_{eff}(t_f)}{G_{eff}(t_0)}\simeq 1.2\times 10^{51}
\end{equation}
Accordingly, the Hawking temperature of such a PBH is $T_{BH}=2.8\times 10^{18}$MeV, which is huge and highly unphysical, since we would have noticed all the heaviest particles coming from such evaporations. Therefore, we may conclude that the second scalar-tensor model probably produces unphysical results when the gravitational memory of critical PBH is taken into account. Note however that in the derivation of the cosmological solution $\phi (t)$, we only took into account the scalar field energy density and pressure and we neglected matter.

In addition, the fact that we found unphysical results, does not necessarily means that the model must be excluded right away. This argument is based on the fact that we assumed that the PBH mass is critical, that is, it evaporates today. However, the initial mass of the black hole may have been larger or smaller in comparison to the critical mass. This means that the PBH may have already evaporated in the past or may evaporate completely in the future. These scenarios bring along more constraints that have to be taken into account, like entropy production bounds, cosmological nucleosynthesis bounds, microwave background spectrum bounds, see for example \cite{barrow1}. In this paper our interest was mainly for the critical mass case.

\subsection{Comparison of Model 1 and Model 2 with Brans-Dicke Models}

In order to have a clear idea on how the scalar-tensor models behave as the time varies, we shall compare the solutions (\ref{phit1}) and (\ref{phit11}), with the ones corresponding to the simplest scalar-tensor theory, namely Brans-Dicke model, with and without cosmological constant. We must have in mind that we did not include any matter content in the scalar-tensor models we used in the previous sections, but we will use matter profiles for the Brans-Dicke model. The general action in the Jordan frame describing a Brans-Dicke model with cosmological constant and matter is:
\begin{align}\label{generaleqnbd}
S=&\int \mathrm{d}x^4\sqrt{-g}\Big{[}\frac{1}{2}\varphi R-\frac{\omega}{\varphi }g^{\mu \nu}\partial_{\mu}\varphi \partial_{\nu}\varphi -\Lambda +L_{matter}\Big{]}
\end{align}
\begin{figure}[h]
\begin{center}
\includegraphics[scale=.8]{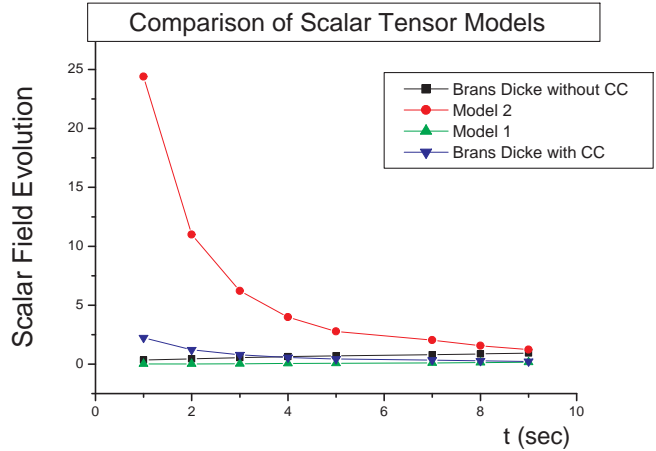}
\end{center}
\caption{Comparison of the evolution of the scalar field $\phi (t)$ as a function of t, for Brans-Dicke models with and without cosmological constant with Model 1 and Model 2. The time axis has been properly rescaled. The same applies for the y-axis.} \label{comparisson}
\end{figure}
By varying with respect to the metric and with respect to the scalar field, we obtain the Einstein equations describing the cosmological evolution of the Brans-Dicke model, which for a flat FRW metric are equal to:
\begin{align}\label{cosmeqnsbdcc}
& 6\varphi H^2=\frac{1}{2}\dot{\phi}+\Lambda+\rho-6H\dot{\varphi} \\ \notag &
\ddot{\varphi}+3H\dot{\varphi}=4\zeta^2\Lambda \\ \notag &
\dot{\rho}+4H\rho =0
\end{align}
where $\varphi=\frac{1}{2}\xi \phi^2$. A particularly simple solution exists if we assume that the matter density is $\rho =\mathrm{const}$ and in addition $H\simeq 0$, which is the following,
\begin{equation}\label{soltion}
\varphi (t)=2\zeta^2\Lambda t^2t^2+\phi_1t+\phi_0
\end{equation}
Hence, the effective gravitational constant is:
\begin{equation}\label{effbdgraconst}
G_{eff}(t)=\frac{1}{16\pi }\frac{1}{2\zeta^2\Lambda t^2+\phi_1+\phi_0}
\end{equation} 
Recall that $G_{eff}(t)\sim \varphi (t)^{-1}$. We shall use the usual values appearing in the literature, in reference to the parameters $\xi ,\Lambda ,\zeta$, with $\zeta^{-2}=6+\epsilon \xi^{-1}$. The comparison of the Brans-Dicke model with cosmological constant with the two scalar models appears in figure 1, where we have properly rescaled the time axis and the $y$-axis. We can work out the Brans-Dicke model without cosmological constant to see how this modifies the field evolution. In this case we shall consider another matter profile, of the form $\rho (t)=\rho_0t^{\beta}$, with $\beta <-2$. The solution $\varphi (t)$ in the Jordan frame reads,
 \begin{equation}\label{soltionbdwcc}
\varphi (t)=\phi_0t^{\beta+2}
\end{equation}
and so the effective gravitational constant is:
\begin{equation}\label{effbdgraconstbdwcc}
G_{eff}(t)=\frac{1}{16\pi }\frac{1}{\phi_0 t^{\beta+2}}
\end{equation}
The comparison of solution (\ref{soltionbdwcc}), with the other three solutions appears also in figure 1.

\section{Possible Perspectives and Implications}

\subsection{A Brief Discussion on Anti-evaporation of PBH in $F(R)$ Gravity, PBH and Supermassive Black Holes}
  
Unlike how ordinary black holes are formed, PBH originate from non-collapsing stellar matter and particularly are formed from primordial density fluctuations. Imagine the evolution of an initial Schwarzschild-de Sitter primordial black hole with it's mass given by the following relation 
\begin{equation}\label{inmasspbh1smbh}
M_i=\Big{(}\frac{t_0}{aG^2(t_0)}\Big{)}^{1/3}\Big{(}\frac{G(t_0))}{G(t_f)}\Big{)}^{2/3}
\end{equation}
with $t_f$ the time at which the black hole was formed and $t_0$ the present age of the universe. In certain $F(R)$ theories there is the possibility of obtaining anti-evaporating black holes \cite{hawking}, see for example \cite{odintsovantievaporation}. Hence, by suitably choosing the parameters, we may start with an initial PBH with mass $M_i$ which will increase in time. This black hole will develop to be a supermassive black hole and instead of being evaporated by it's own Hawking radiation, it will have an exponential increase in it's mass (see also \cite{supermassivepbh}),
\begin{equation}\label{expincreaemass}
M_*(t)=M_ie^{t-t_i/\tau}
\end{equation} 
Notice that the supermassive black hole of this type will carry information related to the time of it's creation, since the initial mass $M_i$ depends on the gravitational constant $G(t_f)$ at the time $t_f$. Work is in progress to find a potential candidate $F(R)$ theory that gives acceptable cosmological features and anti-evaporation of black holes at the same time. Since this task is beyond the scopes of this article, we defer this work to a future article. The possibility that a PBH may develop to be a supermassive black hole was extensively studied in reference \cite{supermassivepbh}.

 \subsection{Antigravity and Gravitational Memory?}
 
As a final task, we shall briefly discuss what happens in the case we take into account gravitational memory in antigravity regimes. Antigravity regions are possible in the context of $F(R)$ theories \cite{odintsov2}. If our universe enters an antigravity phase, then the formation of black holes and other astrophysical objects is not so likely. This is owing to the fact that any particle will repel all other particles and even though local density fluctuations may exist, in the end there is no gravitational attraction. Even if we take into account the speculative scenario that antimatter repels always matter, in an antigravity regime matter repels matter and antimatter. If however antimatter attracts matter in an antigravity phase of the universe and if matter repels matter, maybe this could lead to the formation of heavy astrophysical objects. In addition, a smooth transition from antigravity to gravity could lead to decoupling of matter from antimatter, having in mind that matter and antimatter repel each other in normal gravity phases. If the gravitational objects that were formed at the antigravity phase would have negative Hawking temperature and negative gravitational constant. However, there can be no smooth transition from antigravity to gravity phase, see \cite{odintsov2} and references therein. Particularly, in reference \cite{odintsov2} an antigravity scalar-tensor model of the following form was studied:
\begin{align}\label{generaleqnbdantigrav}
S=&\int \mathrm{d}x^4\sqrt{-g}\Big{[}\frac{1-\varphi^2}{12} R-\frac{1}{2}g^{\mu \nu}\partial_{\mu}\varphi \partial_{\nu}\varphi -J(\varphi )\Big{]}
\end{align}
The corresponding $F(R)$ gravity can be easily obtained using the method of section 1 (see reference \cite{odintsov2}) and is equal to:
\begin{align}\label{generaleqn4}
S=&\int \mathrm{d}x^4\sqrt{-g}F(R),
\end{align}
with $F(R)$ standing for:
\begin{equation}\label{frequation1}
F(R)=\frac{e^{\eta (\varphi (R))}}{12}\Big{(}1-\varphi^2(R)\Big{)}R-e^{2\eta (\varphi (R))}J(\varphi (R))
\end{equation}
In addition, the real function $\eta(\varphi )$ is chosen in such a way so that the following equation is satisfied:
\begin{equation}\label{cnstr}
(1+2\varphi^2)\eta'(\varphi)^2-4\eta'(\varphi )-4=0
\end{equation} 
so that the kinetic term of the scalar field is eliminated. This model clearly provides us with a negative gravitational constant for some values of the scalar field $\varphi $, this however is not a smooth transition indicating a big crunch-big bang transition. What would be interesting is to invert the coupling of the Ricci scalar but this however would imply some sort of instability in the action. Bottom line, if in an antigravity phase of the universe it is possible to have formation of gravitational objects, gravitational memory would have resulted to a negative Hawking temperature for these gravitational objects. But as we mentioned, this scenario is speculative and work is in progress towards this issue.

\section*{Conclusions} 
 
In this paper we studied the effect of gravitational memory on PBH, in the theoretical context of $F(R)$ related scalar-tensor theories. We used a particular form of scalar-tensor theories with a non-minimally coupled scalar field and we demonstrated the equivalence of the scalar-tensor theory with an $F(R)$ theory. Using two characteristic examples, we examined the impact of gravitational memory on PBH that formed sometime in the early universe. We assumed that the mass of the PBH is such that the black hole evaporates at present time and then used the specifics of each model in order to see the implications on black hole observable data. As we evinced, one of the two models produces observable quantities that would not agree with present time observational data. This implies that either the gravitational memory mechanism is un-realistic or the mass of the initial PBH is not the critical one, that is equal to the mass of a PBH that evaporates today. In any case, the gravitational memory provides us with another useful theoretical tool to further constrain the $F(R)$ theories to which the scalar-tensor theories are connected with. The purpose of this article was to point the possibility of existence of gravitational memory and to indicate what would this imply to early universe astrophysical objects. It worths examining viable models to explicitly see what impact would these solutions have, bearing in mind that the gravitational memory hypothesis is valid. In addition, the possibility of the formation of supermassive black holes from PBH which are anti-evaporating black holes of some $F(R)$ theory, should also be extensively discussed. We defer these tasks to a future work. 
 
Before closing, let us comment that it would be particularly interesting to address gravitational memory effects related to $F(R)$ theories with non-minimal curvature-matter coupling \cite{bertolami,bertolami1}, or their equivalent scalar-tensor counterparts. This task could be potentially interesting due to non-trivial quantum effects that the non-trivial matter-coupling of the type $F(R,T)$ can imply, with $T$ the trace of the energy momentum tensor. In addition, the evaporation picture of the PBH may be significantly altered. We hope to address this issue in the near future.

\end{document}